\documentclass[graybox]{svmult}

\usepackage{type1cm}        % activate if the above 3 fonts are
\usepackage{makeidx}         % allows index generation
\usepackage{graphicx}        % standard LaTeX graphics tool
\usepackage{multicol}        % used for the two-column index
\usepackage[bottom]{footmisc}% places footnotes at page bottom
\usepackage{microtype}        % used for the two-column index

\usepackage{newtxtext}       % 
\usepackage{newtxmath}       % selects Times Roman as basic font
\usepackage{units}
\usepackage{paralist}
\usepackage[strings]{underscore}
\usepackage{url}

\graphicspath{{graph/}}
\newcounter{iXOffset}
\newcounter{iYOffset}
\newcounter{iXBlockSize}
\newcounter{iXBlockSizeDiv2}
\newcounter{iYBlockSize}
\newcounter{iYBlockSizeDiv2}
\newcounter{iDistance}

\makeindex             % used for the subject index
\begin{document}

\title*{Audio Content Analysis}
\author{Alexander Lerch}
\institute{Alexander Lerch \at Georgia Institute of Technology, Atlanta, GA, \email{alexander.lerch@gatech.edu}
%\and Name of Second Author \at Name, Address of Institute \email{name@email.address}
}
\maketitle

%\abstract*{Each chapter should be preceded by an abstract (no more than 200 words) that summarizes the content. The abstract will appear \textit{online} at \url{www.SpringerLink.com} and be available with unrestricted access. This allows unregistered users to read the abstract as a teaser for the complete chapter.
%Please use the 'starred' version of the \texttt{abstract} command for typesetting the text of the online abstracts (cf. source file of this chapter template \texttt{abstract}) and include them with the source files of your manuscript. Use the plain \texttt{abstract} command if the abstract is also to appear in the printed version of the book.}

\section{Introduction}
\label{sec:intro}
    Audio signals contain a wealth of information: just by listening to an audio signal, we are able to infer a variety of properties. For example, a speech signal not only transports the textual information, but might also reveal information about the speaker (gender, age, accent, etc.) and the recording environment (e.g., indoors vs.\ outdoors). In a music signal we might identify the instruments playing, the musical structure or musical genre, the melody, harmonies, and tonality, the projected emotion, and characteristics of the performance as well as the proficiency of the performers. An audio signal can contain and transport a wide variety of content beyond these examples; the field of \textit{Audio Content Analysis (ACA)}\index{Audio Content Analysis} aims at creating and using algorithms for automatically extracting this information from the raw (digital) audio signal \cite{lerch_introduction_2012}, enabling us to sort, categorize, segment, and visualize the audio signal based on its content. Use cases include applications such as content-based automatic playlist generation and music recommendation systems, computer-assisted music production and editing, and intelligent music tutoring systems identifying mistakes and areas of improvement for young instrumentalists. 

    This chapter gives an overview of ACA techniques and applications. While the processing of speech signals is covered in Chapter~\ref{X}, this chapter focuses on music signals, where ACA is often referred to as \textit{Music Information Retrieval} (MIR)\index{Music Information Retrieval} \cite{schedl_music_2014,burgoyne_music_2015}, although the latter additionally encompasses the analysis and generation of symbolic (non-audio) music data such as musical scores.%,lerch_music_2014
    
    \subsection{Audio content}
    \label{ssec:content}
        It is important to identify the main sources of content in order to understand what encompasses the content of a signal. In music recordings, the content can be traced back to three origins: 
        \begin{itemize}
            \item   \textit{Composition}: while in western classical music, this would be a written full score, in other musical styles it might be a lead sheet or simply a musical idea. The content related to the composition allows us to recognize different renderings of the same song or symphony as being the same piece. In most genres of western music, this encompasses musical elements such as melody, harmony, and rhythm.
            \item   \textit{Music performance}: the performance realizes the composition in a unique acoustic rendition, the actual music that can be perceived. A performance communicates the explicit information from the composition but also interprets and adds to this information. This happens through variations of tempo, timing, dynamics, and playing techniques.
            \item   \textit{Music production}: since the input of an analysis system is usually an audio recording, the choices made during the recording as well as the editing and processing can significantly influence the final result \cite{maempel_musikaufnahmen_2011}. Microphone positioning, filtering, and editing are examples of the production teams' impact on the final recording.
        \end{itemize}
        An analogy can again be found for speech recordings: the composition corresponds to the text, the performance to the actual speech, and the production to the recording and audio processing.
        
        %Although the term content has been mentioned frequently, a good characterization is still missing. 
        From a musical point of view, audio content can categorized into timing (tempo and rhythm), dynamics, pitch, and timbre. 
        %Table~\ref{tab:content} lists content examples for these categories.
        %\begin{table*}
            %\centering
                %\begin{tabular}{l|ccc}
                            %& Idea                      & Performance                   & Production\\ \hline
                    %Timing  & rhythm                    & tempo, micro-timing           & time-stretching, editing\\
                    %Dynamics& potentially specified   & accents, dynamics variations  & gain, dynamics processing\\
                    %Pitch   & explicitly specified      & vibrato, expressive intonation& pitch-correction/pitch-shifting\\
                    %Timbre  & instrumentation           & playing techniques            & equalization, reverberation
                %\end{tabular}
            %\caption{Audio content examples for different categories}
            %\label{tab:content}
        %\end{table*}
        The combination of specific characteristics and variations across these categories can convey higher level content such as musical genre or conveyed emotion.
        In addition to the musical content alone, a recording might also contain additional content such as song lyrics. 
        %It should be noted that there is also music-related bibliographical information that is not or only implicitly in the audio file, for instance, the publication date.

    \subsection{Generalized audio content analysis system}
    \label{ssec:system}
\begin{figure}%
			\centering
            \includegraphics{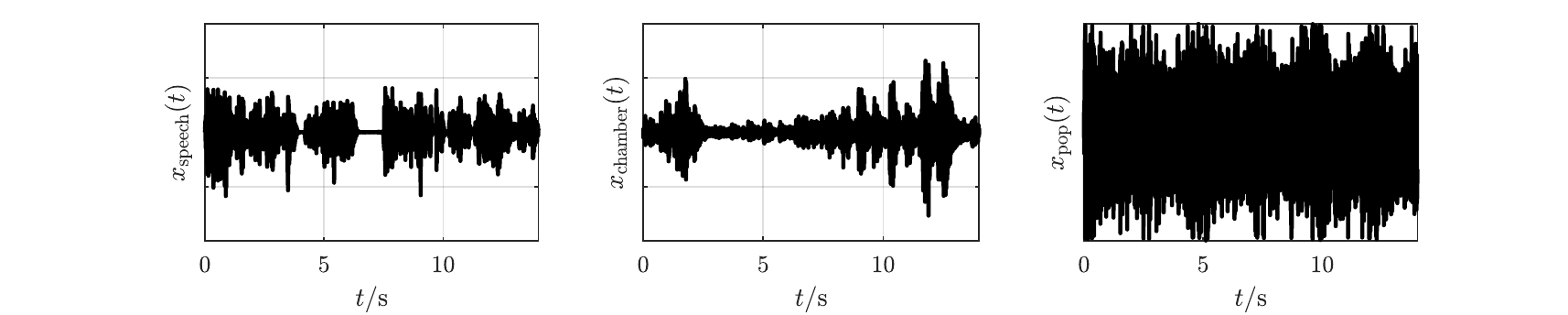}%
            \caption[Audio Waveform]{Waveform representation of three different audio recordings: speech (left), string quartet (middle), pop (right)}%
            \label{fig:waveform}%
        \end{figure}

        A digital audio signal is represented as a series of numbers, so-called samples (see Chap.\ \ref{X}). Direct inspection of these samples~---in case of CD audio quality 44100 per second---~does not necessarily allow the observer to draw conclusions about the audio content. A long term context, however, might still give some insights: Fig.~\ref{fig:waveform} shows the common \textit{waveform}\index{Waveform} representation (sample values over time) of three different audio signals: a string quartet recording, a pop production, and speech. 
        
\begin{figure}
			\centering
			\begin{footnotesize}
    \begin{picture}(85,26)
        \setcounter{iXOffset}{0}
        \setcounter{iYOffset}{5}
        \setcounter{iXBlockSize}{27}
        \setcounter{iYBlockSize}{16}
        \setcounter{iYBlockSizeDiv2}{8}
        \setcounter{iDistance}{7}

        \addtocounter{iYOffset}{\value{iYBlockSizeDiv2}}
        \addtocounter{iYOffset}{-2}

        %\addtocounter{iXOffset}{-1}
        \put(\value{iXOffset}, \value{iYOffset})
            {\text{{\shortstack[c]{Audio\\ Signal}}}}
        \addtocounter{iXOffset}{1}

        \addtocounter{iYOffset}{2}
        \addtocounter{iXOffset}{\value{iDistance}}

        \put(\value{iXOffset}, \value{iYOffset})
            {\vector(1,0){\value{iDistance}}}

        \addtocounter{iXOffset}{\value{iDistance}}
        \addtocounter{iYOffset}{-\value{iYBlockSizeDiv2}}
        
        \put(\value{iXOffset}, \value{iYOffset})
            {\framebox(\value{iXBlockSize}, \value{iYBlockSize}) {{\shortstack[c]{Feature\\ Extraction}}}}

        \addtocounter{iXOffset}{\value{iXBlockSize}}
        \addtocounter{iYOffset}{\value{iYBlockSizeDiv2}}

        \put(\value{iXOffset}, \value{iYOffset})
            {\vector(1,0){\value{iDistance}}}

        \addtocounter{iXOffset}{\value{iDistance}}
        \addtocounter{iYOffset}{-\value{iYBlockSizeDiv2}}

        \put(\value{iXOffset}, \value{iYOffset})
            {\framebox(\value{iXBlockSize}, \value{iYBlockSize}) {{\shortstack[c]{Inference,\\ Decision,\\ Classification}}}}

        \addtocounter{iXOffset}{\value{iXBlockSize}}
        \addtocounter{iYOffset}{\value{iYBlockSizeDiv2}}

        \put(\value{iXOffset}, \value{iYOffset})
            {\vector(1,0){\value{iDistance}}}

        \addtocounter{iXOffset}{\value{iDistance}}
        \addtocounter{iYOffset}{-2}

        \addtocounter{iXOffset}{1}
        \put(\value{iXOffset}, \value{iYOffset})
            {\text{{\shortstack[c]{Meta\\ Data}}}}
        
    \end{picture}
\end{footnotesize}	
			\caption[General Audio Content Analysis System]{General processing stages of a system for audio content analysis}
            \label{fig:ACA_system}	
		\end{figure}
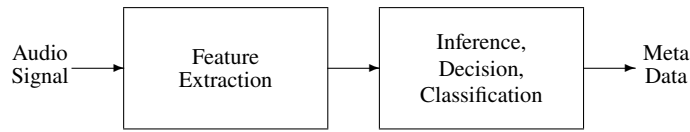

        The general shape of the waveform envelope already enables an observer to differentiate between these different signals by deriving descriptive properties related to dynamic range, fluctuations, and pauses. This can be modeled algorithmically: first, one or more descriptors\index{Descriptor} which capture general properties of the audio signal can be defined (for instance, a measure for how often and how much the waveform envelope changes), and second, a mapping for different descriptor ranges to the audio signal class can be heuristically found (for instance, thresholding a descriptor for classifying the signal as pop). These two processing steps form a simplified model of a generalized system for ACA as shown in Fig.~\ref{fig:ACA_system}: the first stage extracts descriptors, also commonly referred to as \textit{audio features}\index{Audio Feature}%\index{Feature!Audio}
        , and the second stage infers the desired content information from the features. There are some parallels of these two stages to two steps in human information processing, namely perception and cognition.
    
        \subsubsection{Feature extraction}
        \label{sssec:feature}

            The \textit{Feature Extraction} stage has two main objectives: first, it reduces the overall amount of data to be processed, leading to a compact representation of the content. For example, some audio classification systems use only dozens or hundreds of feature values to describe a complete song instead of millions of samples. Second, it focuses on the relevant aspects of the content, stripping away irrelevant and possibly redundant information. If, for example, a system is supposed to detect the fundamental frequencies of an audio signal, the feature set should probably be invariant to loudness and timbre. It is important to note that features do not necessarily have to be musically meaningful or even interpretable by humans, it suffices that they contain the relevant information needed for inference to provide a correct mapping to the human-understandable target meta data. This is particularly true for most state-of-the-art systems: while for a long time experts carefully designed features capturing specific content information (such as the envelope variation in the simple example above), the last decade has seen a new generation of data-driven systems which automatically learn such features from data \cite{humphrey_feature_2013}. Prominent examples of these modern systems are neural networks which can automatically learn a compressed representation of the input data as features.
            
            The input of the majority of ACA systems is either the waveform as discussed above or some kind of spectrogram representation. A spectrogram\index{Spectrogram} is a pseudo-3D plot that, compared to the waveform, often gives more insights into the frequency content of the signal by plotting the magnitude of many overlapping Short Time Fourier Transforms (STFT)\index{Short Time Fourier Transform} over time. 
        
        \begin{figure}%
			\centering
            \includegraphics{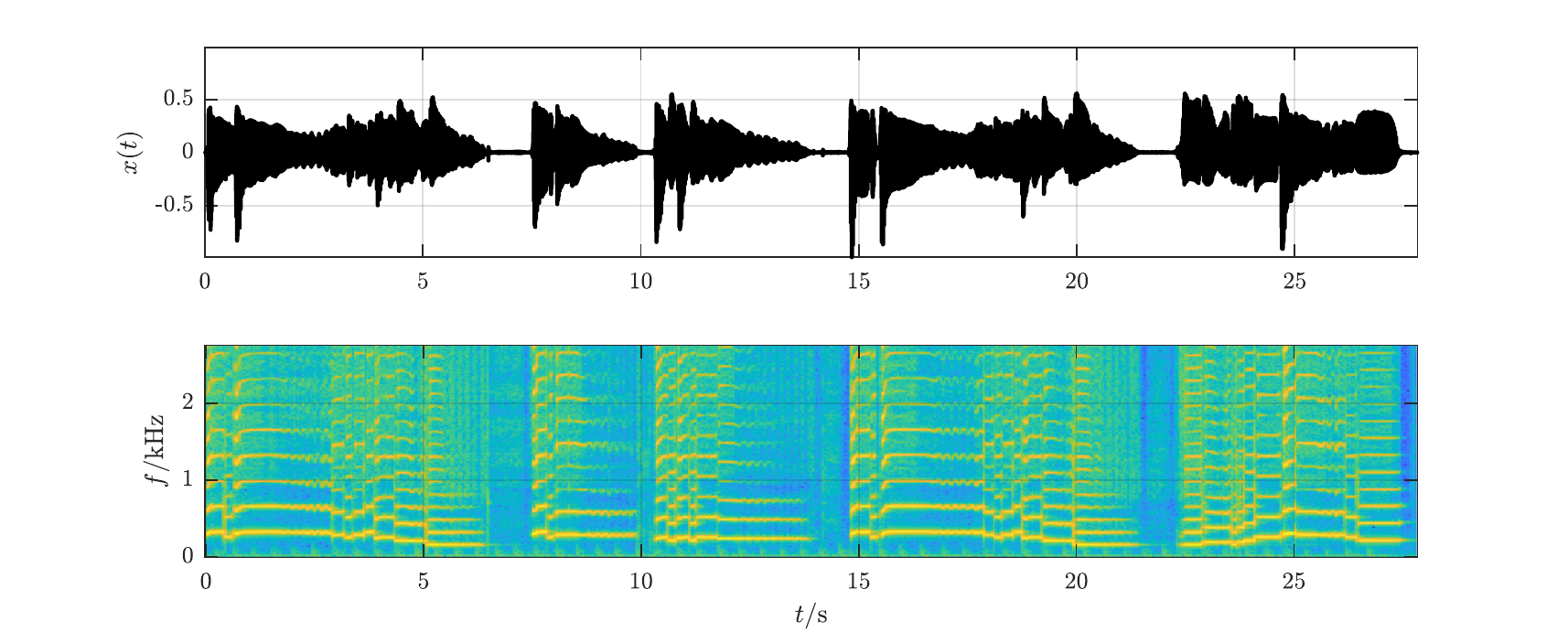}%
            \caption[Audio Spectrogram]{Waveform (top) and spectrogram (bottom) view of a single-voiced saxophone recording of the jazz standard 'Summertime'}%
            \label{fig:specgram}%
        \end{figure}
        
            Figure~\ref{fig:specgram} shows the first 24 bars of a single saxophone playing the jazz standard 'Summertime;' the top visualizes the common waveform representation (x-axis: time, y-axis: amplitude) and the bottom displays the spectrogram (x-axis: time, y-axis: frequency, color: amplitude). Each column of the spectrogram is one magnitude spectrum of a short block of samples with low values colored blue and high values colored yellow. While the phrases are easily identifiable in both representations, the spectrogram also shows that
            \begin{inparaenum}[(i)]
                \item   it is a recording of a single monophonic instrument (clear spectral structure of fundamental frequency and harmonics at integer multiples),
                \item   it is an instrument that allows vibrato (see seconds $3$ and $18$),
                \item   it is an instrument with a significant number of harmonics (number and strength of ``parallel'' lines), and that
                \item   the melody can be directly derived from the visualization by identifying the lowest frequency of the harmonic series and mapping it to musical pitch.
            \end{inparaenum}
        
        \subsubsection{Inference stage}
        \label{sssec:inference}
        
            The second stage of the ACA system, the \textit{inference}\index{Inference}, takes the extracted features and maps them into a domain both usable and comprehensible by humans. In other words, it interprets the feature data and maps it to meaningful meta-data, such as a class label for the example in Fig.~\ref{fig:waveform} (speech, chamber music, pop) or the pitch of a melody. This inference system can be an expert-defined algorithm, a classifier, or a regression algorithm. The more condensed and the more meaningful the features are, the less powerful the inference algorithm has to be and vice versa: a raw feature representation close to the audio sample requires a sophisticated inference approach. 
            
            It is important to note that no machine learning system will work with hundred percent accuracy in all but the most trivial tasks. Every machine model of data and its patterns will always be imperfect just like a human annotating data, and it is the goal of the scientists and engineers to minimize the number of errors.

%\begin{itemize}
	%\item   task definition and terminology: here music
    %\item   applications and use cases
    %\item   content to be extracted
    %\item   general content analysis system
    %\item   historic overview: signal processing driven/rule based, feature design, end-to-end systems
    %\item   chapter overview
%\end{itemize}

\section{Music transcription}
\label{sec:transcription}

    \textit{Music transcription}\index{Music transcription} systems estimate (explicit or implicit) score information from the audio recording. Music transcription broadly defined encompasses a variety of extraction tasks covering, for instance, melody, chords, musical key, musical instruments, rhythm, time signature, and structure. 
    %This chapter will present a few representative examples starting with the introduction of methods nowadays considered basic before ending with a glimpse into modern, state-of-the-art approaches.
    The basics of music transcription systems will be explained here by introducing simple example approaches to various tasks.
    
    \subsection{Musical key detection}
    \label{ssec:key}
        Estimating the key of a musical audio signal has multiple applications ranging from large-scale musicological studies to enabling DJs to automatically identify songs with tonal compatibility for mash-ups. Key detection systems\index{Musical Key Detection} could work very reliably if they could simply utilize a transcribed version of all notes with pitch and length as the key is mostly defined by the pitch content of a piece. Practically, however, this approach is not necessarily the most robust approach as, on the one hand, the transcription of pitches from polyphonic audio remains to be challenging and  error prone and, on the other hand, key detection approaches can work reasonably well without requiring such detailed information \cite{pauws_musical_2004,chuan_polyphonic_2005,izmirli_template_2005}.
        
        \subsubsection{Pitch chroma}
        \label{ssec:chroma}
\begin{figure}%
                \centering
                \includegraphics{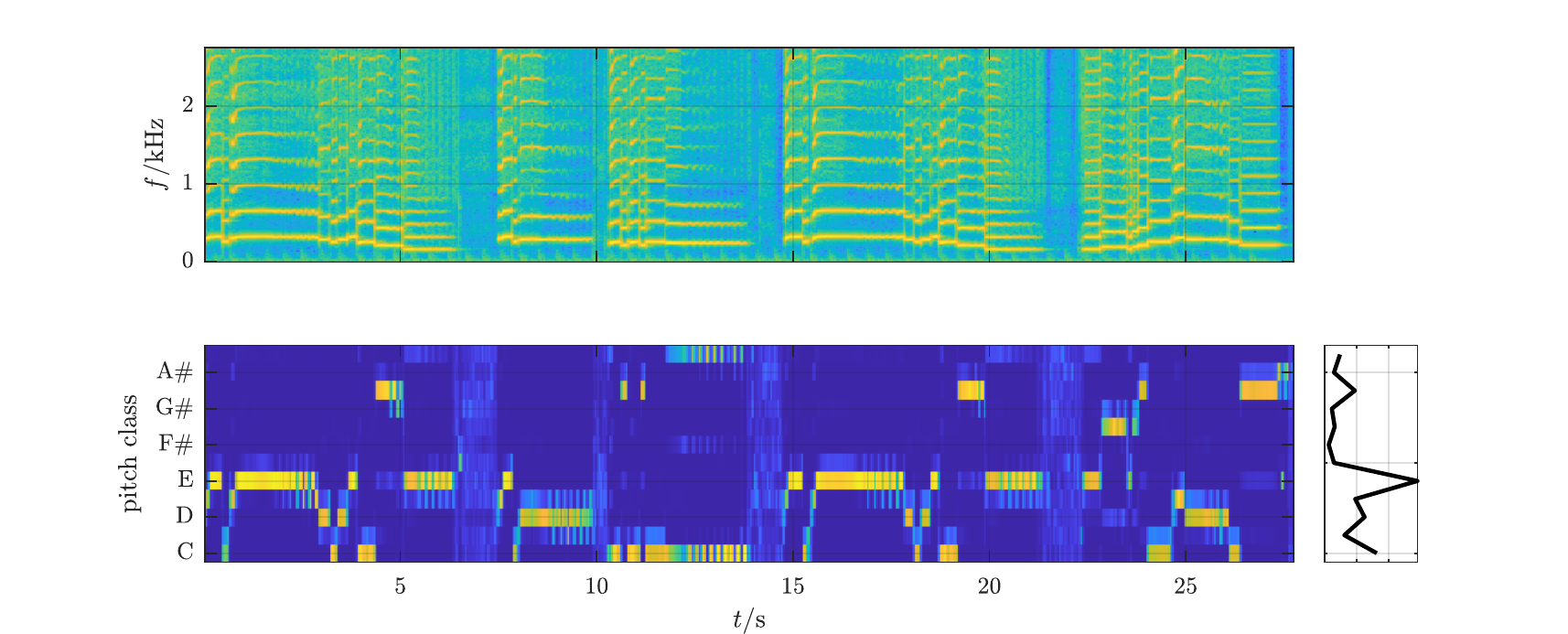}%
                \caption[Pitch chroma]{Spectrogram (top) and pitch chroma (bottom) view of a single-voiced saxophone recording of the jazz standard 'Summertime'}%
                \label{fig:pitchchroma}%
            \end{figure}

            The most common feature that is used for key detection is the so-called average \textit{pitch chroma}\index{Pitch chroma}, which is an easy-to-extract octave-independent approximation of the pitch content in the audio. Its use was first proposed in the context of chord detection \cite{fujishima_realtime_1999}. Similar to many other low-level features, the pitch chroma is usually derived from a STFT computed on the blocked audio signal. 
            %The spectrogram in Fig.~\ref{fig:specgram} shows such a series of STFTs as each column corresponds to the magnitude spectrum of a block of audio samples at that specific time, and each single STFT indicates the magnitude of a specific frequency. 
            For pitch chroma computation, the frequency bins are grouped to the frequency of the closest musical pitch, e.g., A4. Then, all the magnitudes belonging to the same pitch class (e.g., A3, A4, A5) are accumulated over the octaves, resulting in a 12-dimensional pitch class vector (C, C\#, D, D\#, etc.) per audio block. The result is shown in Fig.~\ref{fig:pitchchroma} (bottom); it is similar to the spectrogram in the sense that it is a pseudo-3D plot with time on the x-axis, however, the y-axis represents the 12 pitch classes C--B now.
            
            The pitch chroma is a fitting feature for detecting the musical key as it focuses on the tonal content, reduces timbre and rhythm interference, and removes key-irrelevant octave information \cite{gomez_tonal_2006,muller_information_2007}. 
            
        \subsubsection{Inference}
        \label{ssec:inference}
\begin{figure}%
			\centering
            \includegraphics{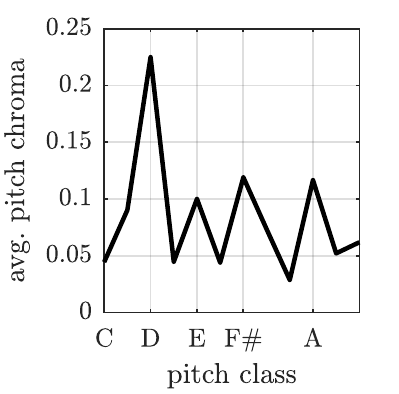}%
            \caption[Average Pitch Chroma]{Average pitch chroma of a pop song in the key of D Major}%
            \label{fig:avgpitchchroma}%
        \end{figure}

            Under the simplifying assumption that the key will not change over the course of the piece of music (an assumption mostly valid for some genres such as rock and pop but invalid for others, e.g., classical music), the average pitch chroma of the whole piece gives a good approximation of the overall pitch content of the piece. Figure~\ref{fig:avgpitchchroma} displays an average pitch chroma extracted from a pop song in key D Major; it shows that the pitches D and A are the most prominent (salient and often occurring) and that unlikely pitch classes such as G\# are less salient.
    
\begin{figure}%
			\centering
            \includegraphics{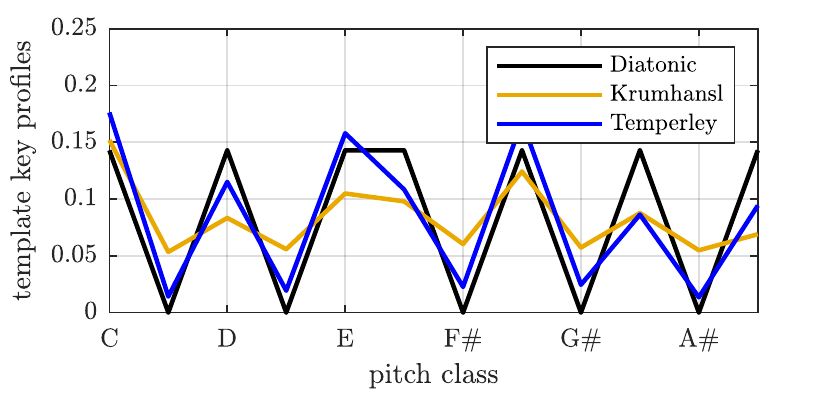}%
            \caption[Template Key Profiles]{Three common template normalized key profiles (C major)}%
            \label{fig:keytemplates}%
        \end{figure}

            A simple key detection system uses such an extracted pitch chroma and infers the estimated key by comparing it to previously defined pitch class distribution templates, also referred to as key profiles. Examples for such templates as shown in Fig.~\ref{fig:keytemplates} can be derived from music knowledge (diatonic), listening experiments on tonality (Krumhansl) \cite{krumhansl_cognitive_1990}, or data (Temperley) \cite{temperley_tonal_2007}. One disadvantage of the diatonic profile is that the profile is identical between a major key and its relative (aeolic) minor key; the other two key profiles solve this issue by allowing different weights for different scale degrees.
    
            As the key profile of, for instance, C\# Major can be assumed to be identical to the C Major key profile but shifted by one semi-tone, only two 12-dimensional templates are required for the basic task of detecting major and minor keys. The final key estimate is the key that minimizes the distance between the extracted average pitch chroma and the (shifted) major and minor key profile templates.
            
        \subsubsection{Results}
        \label{ssec:keyres}
            Modern key detection systems achieve correct detection rates of approximately \unit[60--80]{\%}, depending on the data used for evaluation. The most common errors are confusions with the relative key (e.g., G Major vs.\ E Minor), the closely related keys (e.g., D Major vs.\ A Major), and the parallel key (e.g., A Major vs.\ A Minor),\footnote{cf.~\url{https://www.music-ir.org/mirex/wiki/2019:Audio_Key_Detection_Results}, last accessed 01/14/2020} which is easily understandable due to the significantly overlapping pitch content.

    \subsection{Monophonic fundamental frequency detection}
    \label{ssec:f0}
        The estimation of the varying fundamental frequency $f_0$ from a single-voiced audio recording  enables a variety of different applications ranging from pitch correction over automatic accompaniment systems to karaoke systems. It is generally considered to be a solved problem, although some standard systems sometimes still lack the robustness required by specific applications. Fundamental frequency detection\index{Fundamental frequency detection!Monophonic} systems detect repeating patterns in tonal, quasi-periodic components of the signal. Most approaches build on the basic assumption that the single-voiced signal is a weighted superposition of multiple sinusoidals with frequencies at integer multiples of the fundamental frequency. Once the fundamental frequency is detected, it can be easily mapped to a musical pitch.
        
        One of the most intuitive ways of approaching fundamental frequency detection is to measure the distance between zero-crossings and local maxima (or minima) to estimate the fundamental period length, the inversion of which is the fundamental frequency $f_0$. While this is a simple way of approaching this problem, the results are too unreliable to make it practically usable, especially in the case of numerous harmonics.
        
        Over the past few decades, a variety of approaches to fundamental frequency detection have been proposed; the following sections present representative approaches for two analysis domains, the time domain and the frequency domain.
        
        \subsubsection{Auto correlation function}
        \label{ssec:acf}
\begin{figure}%
			\centering
            \includegraphics{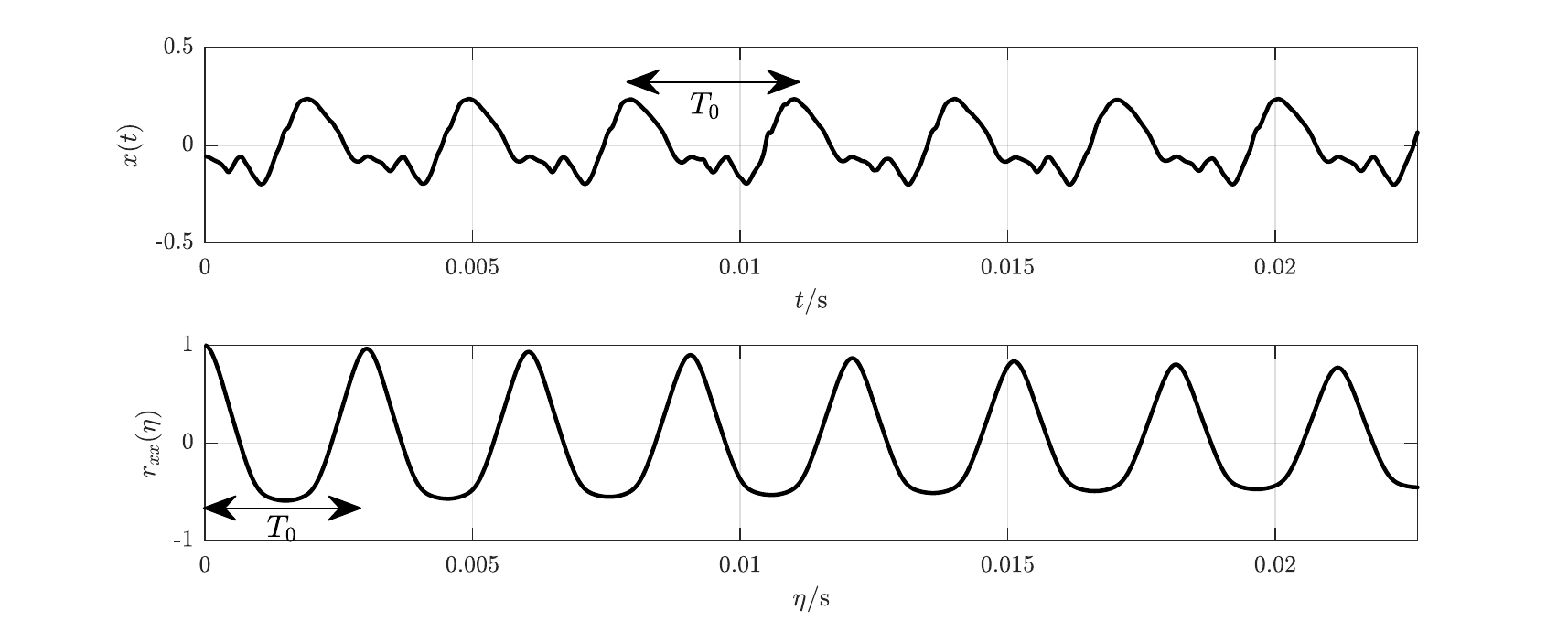}%
            \caption[Auto correlation function]{Normalized auto correlation function (bottom) of a periodic audio input (top)}%
            \label{fig:acf}%
        \end{figure}

            Nearly every pitched signal is periodic with its fundamental period length. An established way of detecting this periodicity or self-similarity is the \textit{Auto Correlation Function (ACF)}\index{Auto Correlation Function}. Assuming that the fundamental frequency of a signal does not significantly change within a short block of samples of length $\mathcal{N}$, the periodicity can be found by multiplying this block of samples $x(n)$ with a shifted version of itself $x(n+\eta)$ and summing the result for each $\eta$:
            \begin{equation}
                r_\mathrm{xx}(\eta) = \sum\limits_{n=0}^{\mathcal{N}}{x(n)\cdot x(n+\eta)}.
            \label{eq:acf}
            \end{equation}
            The result will be maximal at a shift of $\eta=0$ (maximum self similarity), but it will also show local maxima at multiples of the fundamental period length. This is due to the high similarity of neighboring periods of the periodic signal. Thus, the ACF indicates the similarity per shift $\eta$. The shift of the local maximum is therefore an indicator of the length of the fundamental period length in samples. The ACF is often normalized so that $r_{xx}(0) = 1$. Figure~\ref{fig:acf} visualizes this function (bottom) for an example signal (top) for $\eta \geq 0$.
            
        The ACF has been a popular pitch detection algorithm for many decades \cite{rabiner_use_1977} and related and modified algorithms are still used \cite{mauch_pyin_2014}.

        \subsubsection{Harmonic product spectrum}
        \label{ssec:hps}
\begin{figure}%
			\centering
            \includegraphics{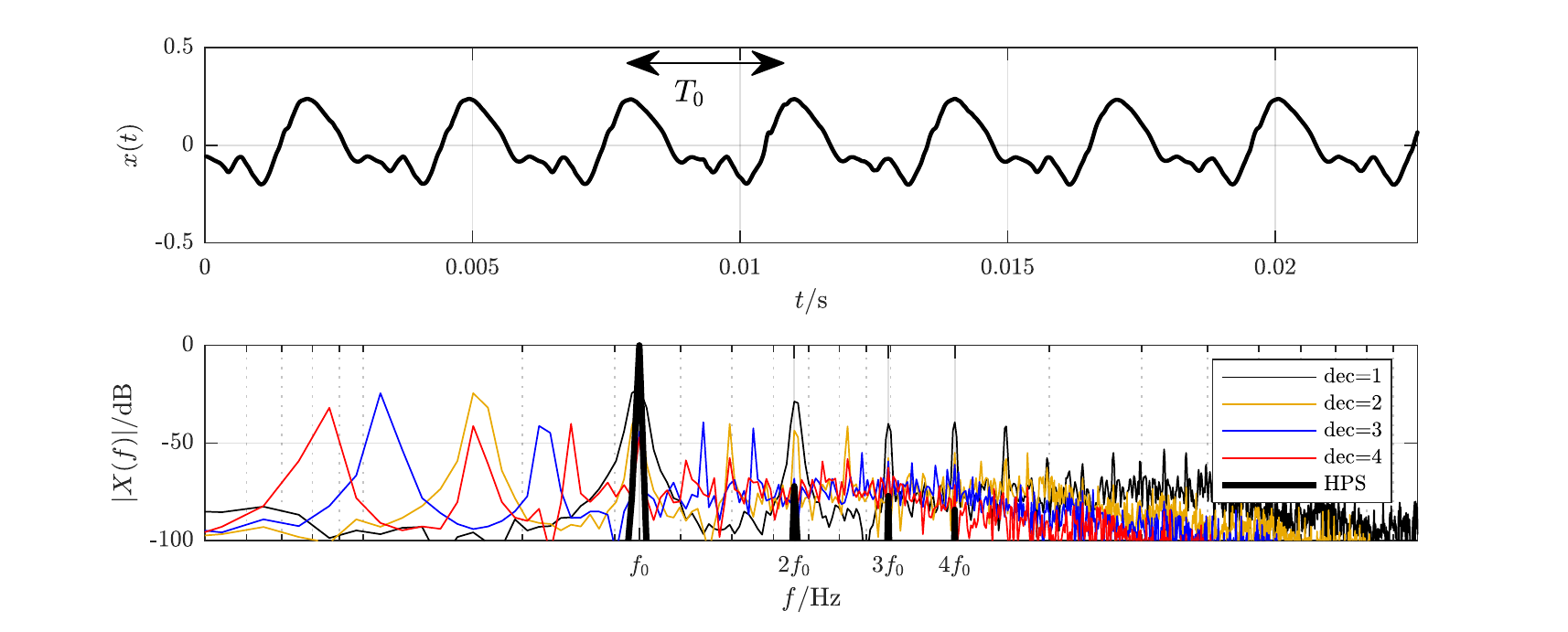}%
            \caption[Harmonic Product Spectrum]{4th order Harmonic Product Spectrum and decimated spectra (bottom) for the signal shown top}%
            \label{fig:hps}%
        \end{figure}

            Intuitively, the frequency domain is a fitting domain to find the fundamental frequency as it is a relatively sparse representation of the frequency content of each block. Simply picking the location of the maximum of the STFT, however, does not lead to a reliable fundamental frequency estimate, as the loudest tonal component of natural sounds is often one of the higher harmonics instead of the fundamental frequency itself. The \textit{Harmonic Product Spectrum (HPS)}\index{Harmonic Product Spectrum} is a way to address this challenge by taking advantage of the comb-like structure of the spectrum of a periodic sound in the frequency domain \cite{noll_pitch_1969}, as the magnitude spectrum of a periodic sound will show local maxima at the location of fundamental frequency as well as its integer multiples. The HPS is computed iteratively: first, the magnitude spectrum is decimated by keeping only every second value so that the length of the spectrum is halved. Whatever the fundamental frequency is, the location of the second harmonic now has the same index as the fundamental frequency of the original spectrum. Multiplying those two spectra will increase the value at the fundamental frequency (multiplication of local maxima) and minimize at other locations of the spectrum (lower magnitudes). This process is iteratively repeated while decimating the spectrum by factors of $3$, $4$, $5$, etc., with the maximum at the fundamental frequency getting more and more pronounced as higher harmonics are multiplied. The location of the maximum of this HPS is, then, the estimated fundamental frequency. Figure~\ref{fig:hps} illustrates the decimated spectra and the resulting HPS. 
            
            The HPS is a simple way of detecting the fundamental frequency in the frequency domain, but it faces some inherent challenges. First, the spectral frequency resolution is often insufficient especially for signals at low frequencies. Second, it might fail for signals in which one harmonic has low magnitude. Third, fundamental frequencies that fall between bin frequencies might not be detected. But even given these issues, the HPS is a clever and intriguing way of using the harmonic structure in the frequency domain to detect the fundamental frequency.
        
    \subsection{Polyphonic fundamental frequency detection}
    \label{ssec:polyf0}
        While fundamental frequency detection for single-voiced recordings is considered a solved problem, a multi-voiced, polyphonic signal\index{Fundamental frequency detection!Polyphonic} still poses challenges to state-of-the-art systems \cite{benetos_automatic_2013}. Historically, this task was initially approached with methods that can be summarized under the term \textit{Iterative Subtraction}: first, a monophonic pitch detector is applied to the signal to detect the most salient fundamental frequency. Second, this frequency is stored is a candidate and all its harmonics are removed from the signal. Third, this process is repeated until the termination criteria are met. Termination criteria can include, for example, a maximum number of voices or a minimum remaining energy in the signal. Approaches utilizing iterative subtraction have been reasonably successful in the 2000s and have been applied to both time domain signals \cite{cheveigne_multiple_1999} and frequency domain signals \cite{klapuri_multiple_2003}.
        
        Later, Non-Negative Matrix Factorization (NMF) became widely-used for polyphonic fundamental frequency detection \cite{smaragdis_non-negative_2003}, and was especially successful  when applied to instruments with quantized frequencies such as the piano \cite{bertin_blind_2007}. NMF attempts to approximate one positive matrix (usually the spectrogram) through the multiplication of two matrices which are randomly initialized and iteratively updated, based on the distance between the target spectrogram and the result of the multiplication. After convergence, the two matrices can be interpreted as the template dictionary, containing all basic sound components that might contribute to the final spectrogram, and the activation, indicating how strong each template is at each time. In the case of the piano, the templates would be all the different piano pitches and the activations indicate the volume of each pitch over time.
        
        Unsurprisingly, the majority of modern systems for polyphonic pitch detection is based on deep neural networks \cite{benetos_automatic_2019}. Most of them work with a spectrogram-like input representation. The achieved accuracies vary, dependent on the data set used for evaluation, between 40 and 70\%. These and more results\footnote{cf.~\url{https://www.music-ir.org/mirex/wiki/2019:Multiple_Fundamental_Frequency_Estimation_\%26_Tracking_Results_-_MIREX_Dataset}, last accessed 01/14/2020} can be found on the website for the so-called MIREX (Music Information Retrieval Evaluation eXchange)\index{MIREX}, an annual event comparing the performance state-of-the-art systems for a variety of Music Information Retrieval tasks on a variety of data and metrics.
        
    \subsection{Musical structure identification}
    \label{ssec:structure}
\begin{figure}%
			\centering
            \includegraphics{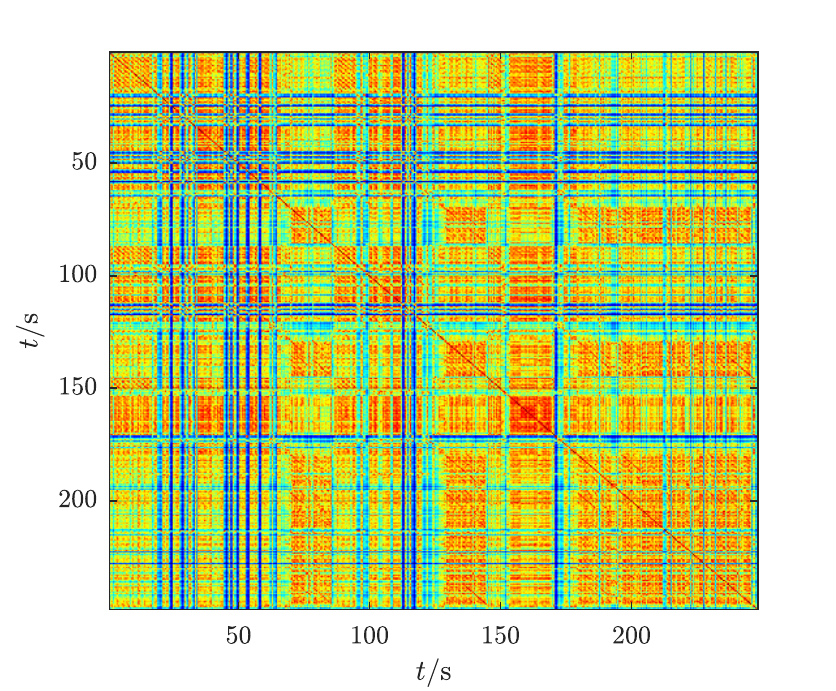}%
            \caption[Self-Similarity Matrix]{Self-Similarity Matrix for Michael Jackson's ``Bad''}%
            \label{fig:ssm}%
        \end{figure}

        Most music is inherently formally organized and structured into various hierarchical levels, starting from groups of notes, bars, and phrases to sections and movements. 
        The detection of musical structure can enable intelligent listening applications, capable of jumping to specific segments such as chorus or verse, or large-scale musicological analyses. 
        The way humans infer structure is based on three main characteristics \cite{paulus_state_2010},
        \begin{inparaenum}[(i)]
            \item   novelty and contrast, meaning something new and possibly unexpected happens,
            \item   homogeneity, meaning that similar parts tend to be grouped together, and
            \item   repetition, meaning that the recognition of a repeated segment indicates a structural segment.
        \end{inparaenum}
        All of these characteristics can be represented through a variety of musical elements including but not limited to harmony, rhythm, melody, instrumentation, tempo, and dynamics. Therefore, a system for automatic structure detection will extract one or more features representing these musical elements and compute an intermediate representation of the feature data. The most common intermediate representation for a structure detection system is a so-called \textit{Self-Similarity Matrix (SSM)}\index{Self-Similarity Matrix}. The SSM is an intuitive way of indicating and visualizing structure. It is computed by extracting a meaningful feature to investigate, for example, the pitch chroma for tonal content. Then, a pairwise similarity is computed between each short-time pitch chroma and all others, leading to a self-similarity matrix as shown in Fig.~\ref{fig:ssm}.
        The resulting SSM is symmetric; both axes of the SSM indicate time and the similarity between $(t_1,t_2)$ equals the similarity between $(t_2,t_1)$. The diagonal indicates the maximum of self-similarity (red). Blocks of constant red color indicate areas of high homogeneity such a held chord or a short repeated pattern, and blue vertical or horizontal lines indicate low similarity to all other points (often rests and pauses). 
        
        The SSM shown in Fig.~\ref{fig:ssm} was computed from Michael Jackson's pop song ``Bad.'' We can clearly identify several structural elements of the song:
        \begin{inparaitem}[]
            \item   the intro stops at about \unit[19]{s},
            \item   the bridge at \unit[60--69]{s} is followed by a chorus (\unit[69--86]{s}) and the same constellation is repeated (as indicated by the line parallel to the diagonal) at \unit[120]{s},
            \item   the high homogeneity of the instrumental is indicated by the red box from \unit[145--170]{s}, and
            \item   the songs ends with four repetitions of the chorus starting at \unit[180]{s} visualized by the high similarity with diagonal structure in the end.
        \end{inparaitem}
        
%0.000	0.365		silence
%0.365	18.825		intro
%18.825	35.648		verse
%35.648	44.059		breaka
%44.059	60.883		verse
%60.883	69.295		bridge
%69.295	86.118		refrain
%86.118	94.529		breakb
%94.529	119.765		verse
%119.765	128.177		birdge
%128.177	144.999		refrain
%144.999	170.233		instrumental
%170.233	178.641		bridge
%178.641	195.463		refrain
%195.463	212.286		refrain
%212.286	229.108		refrain
%229.108	246.695		refrain_(with_complete_ending)
%246.695	246.693		silence

        There are several ways to use the SSM to algorithmically infer the musical structure. One way is to use image processing techniques. For instance, the start of a new segment might be indicated by peaks in the output of a high pass filter (checker kernel) swiped along the diagonal \cite{foote_automatic_2000}. Repetitions, indicated by lines parallel to the diagonal, might be detected with edge detection techniques on the SSM \cite{dannenberg_music_2008}. 
        
        The formal evaluation of structure detection systems remains a challenge: human annotators of musical structure, even if in agreement, tend to annotate different structural levels. For instance, what might just be the two segment structure \mbox{(A) (B)} for one annotator could easily be annotated by another as \mbox{(a b) (c d)}, which makes an evaluation by simply matching the ground truth impractical, given that both annotations are correct and the system might detect either one.

    %\subsection{Musical instrument recognition}
    %\label{ssec:instrument}
    %\subsection{Chord detection}
    %\subsection{Time signature detection}
    %\subsection{Automatic accompaniment}

\section{Music performance analysis and assessment}
\label{sec:mpa}
    Since the ultimate goal of most music transcription tasks is the extraction of a score-like description from an audio file, the performance information is often discarded. There is, however, a notable difference between the two renditions of the same sonata from two pianists or the realization of a jazz standard by different performers \cite{palmer_music_1997}. \textit{Music Performance Analysis (MPA)}\index{Music performance analysis} deals with extracting this performance information from audio recordings. What constitutes performance information varies from genre to genre, but the performance parameters usually considered to be most impactful are dynamics and tempo \cite{lerch_software-based_2009}, although expressive variations of pitch (vibrato, intonation) and playing techniques can play important roles as well.

    \subsection{Dynamics}
    \label{ssec:dynamics}
        Musical dynamics are closely related to loudness, although the absolute perceived loudness is not the only cue to indicate musical dynamics \cite{weinzierl_sound_2018}. A \textsl{forte} passage on a recording is, for example, easily identifiable even at low reproduction volume. Therefore, measures of acoustic intensity are outperformed by humans judging musical dynamics \cite{nakamura_communication_1987}. Even so, it is a valid simplification to reduce the extraction of musical dynamics to extracting simple features representing the energy of a signal as described in Sect.~\ref{ssec:features}.
        
    \subsection{Tempo and beat detection}
    \label{ssec:tempo}
        The tempo of a performance has been shown to convey structural information of the music to the listener \cite{palmer_mapping_1989}. Moreover, tempo and its variation has been linked to the perception of projected emotion \cite{juslin_cue_2000}.
        The tempo of a piece of music is set by a train of quasi-equidistant pulses, so the so-called tactus \cite{lerdahl_generative_1983}. The tactus indicates the beats at which listeners will clap or tap their foot with the music. It is important to note that the tactus is a perceptual concept: while it is determined by groups and accents of note events, the pulse of a tactus may or may not fall on a note event, and a note event may or may not fall on a pulse. The frequency of a tactus is usually in the general range of \unit[1--3]{Hz}, corresponding to \unit[60--180]{BPM} (Beats Per Minute) \cite{fraisse_time_1978}.
        
        Automatically extracting the tempo and the beats from audio recordings enables applications in music recommendation and systems for playlist generation as well as in creative usages such as DJ software.
        
        As the pulse train is periodically repeating, the concept of detecting the tempo shows some similarity to the task of fundamental frequency detection as both tasks focus on estimating the period length of a periodic input signal. The main difference is the time scale of this detection: while for fundamental frequency detection the period lengths of interest range from approximately \unit[0.3--30]{ms}, a typical beat period length is approximately in the range of \unit[300--1000]{ms}.
            
            \subsubsection{Novelty function}
            \label{sssec:novelty}
                Due to these different time scales and because the series of samples carries a large amount of unrelated information, a beat analysis is usually applied to an intermediary representation, the so-called \textit{novelty function} \cite{lerch_introduction_2012}. The novelty function\index{Novelty function} is a time-series of values that has local maxima at positions where ``something new is happening'' such as the onset of a new note or a drum hit and is low at times when nothing new begins such as during a held note.
                While early systems attempted to extract this novelty function from the time domain envelope itself \cite{schloss_automatic_1985}, it can be extracted from a variety of representations or features, including tonal representations such as the pitch chroma and low level features describing spectral shape such as MFCCs or the spectral centroid (see Sect.~\ref{ssec:features}) \cite{lykartsis_beat_2015}.
                The common process for extracting the novelty function involves
                \begin{inparaenum}[(i)]
                    \item   extracting the feature,
                    \item   computing the derivative over time,
                    \item   truncating negative values to zero, and
                    \item   smoothing the result with a lowpass filter.
                \end{inparaenum}
                
\begin{figure}%
			\centering
            \includegraphics{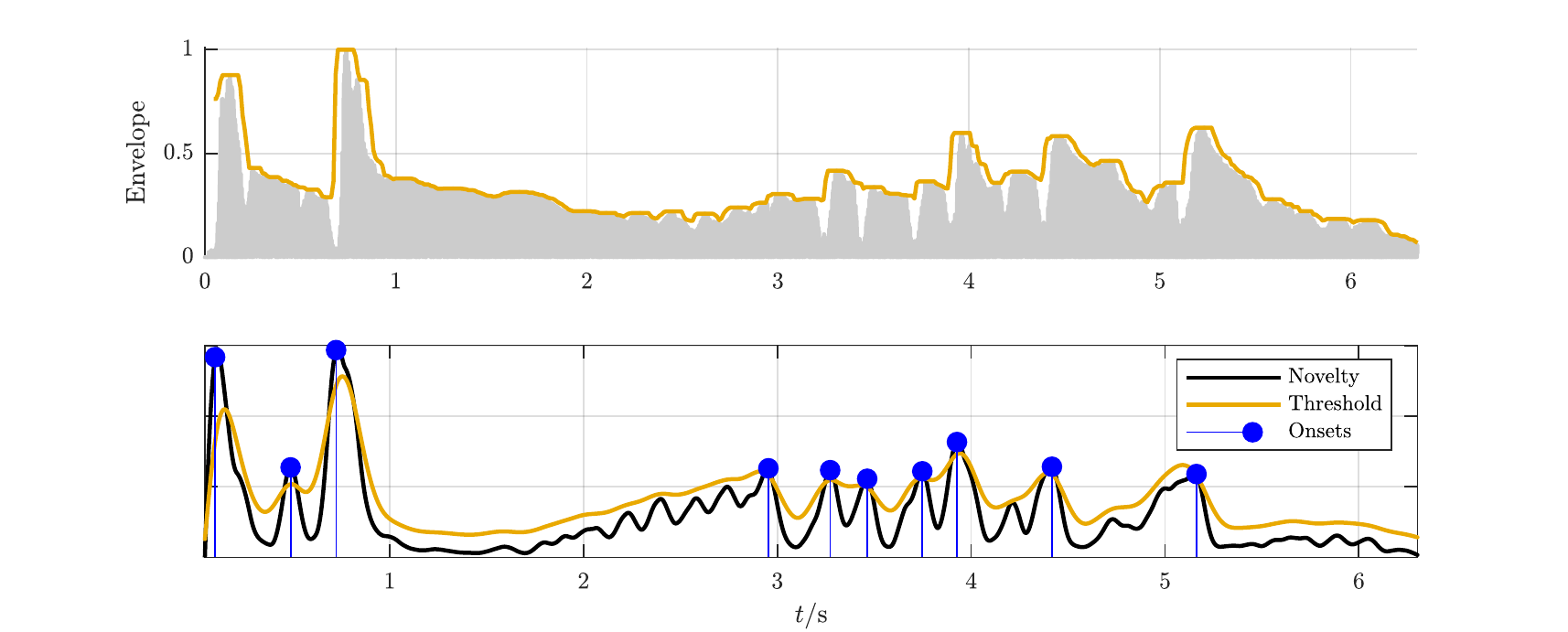}%
            \caption[Novelty function]{Rectified audio envelope (top) and novelty function with picked onsets (bottom)}%
            \label{fig:novelty}%
        \end{figure}

                If the extracted novelty function works as intended, the local maxima indicate note onsets; therefore, picking the peaks of this function is referred to as \textit{onset detection}\index{Onset detection} \cite{bello_tutorial_2005}. Figure~\ref{fig:novelty} shows an example of a novelty function and picked onsets.

            \subsubsection{Tempo induction}
            \label{sssec:tempo}
                As the novelty function gives an indication of note events or other musical events, there is no direct mapping from these events to tempo since, as mentioned above, note events do not necessarily fall on beat events and vice versa. However, either (or both) the novelty function or the series of detected onsets are useful representations for infering the tempo. More specifically, the periodicity of the novelty function allows to estimate the tempo.\index{Tempo Induction}
                
                To give an example of an early system for tempo induction, Scheirer proposed to use a bank of resonance filters \cite{scheirer_tempo_1998}. Each filter is tuned to one possible tempo, for example, \unit[120]{BPM}, and the resonance frequency of the filter with the highest output energy is the most likely tempo. This means that the real tempo can only be detected if it is close to a filter frequency; thus, the number of filters combined with the overall range indicate the possible tempo resolution.
                
                Other simple ways of detecting the tempo include an Autocorrelation analysis of the novelty function \cite{gouyon_beat_2003} or a picking the maximum of the Inter-Onset-Interval histogram \cite{dixon_beat_1999}.
                
                An example for a state of the art system for tempo induction is based on recurrent neural networks for analysis and yielding results in the range of \unit[50--90]{\%} depending on the dataset \cite{bock_accurate_2015}.

            \subsubsection{Beat detection}
            \label{ssec:beat}
                The knowledge of the tempo indicates the period length of the ``foot-tapping rate'' or the distance between the pulses, respectively, however, it does not imply the actual beat locations, sometimes referred to as beat phase. Thus, \textit{beat detection}\index{Beat detection} systems aim at detecting the beat locations from the novelty function. 
                
                One of the first systems proposed to detect beats based on oscillators with adaptive parameters spawned a whole class of beat tracking systems: here, a pulse generator predicts beat locations which are then compared to actual onset times and strengths; dependent on the distance between onset and beat as well as their positions, the pulse generator parameters are adapted to optimize the estimated fit with future onsets \cite{large_beat_1995}. Both beat phase and beat distance are adapted and estimated simultaneously.
                The advantage of these oscillator-based systems is that they are capable of real-time processing; their main disadvantage is slow adaptation to sudden tempo changes.

    %\subsection{Other performance parameters}
    %\label{ssec:performance}
        %\begin{itemize}
            %\item   pitch
            %\item   dynamics
            %\item   playing techniques
            %\item   
        %\end{itemize}
 
    \subsection{Performance assessment}
    \label{ssec:assessment}
        Instead of extracting individual performance parameters, the goal of performance assessment is the estimation of overall ratings for a performance, taking into account parameters spanning the domains pitch (vibrato rates, intonation, tuning and temperament), dynamics (accents, tension), and timbre (playing techniques, instrument settings). This is a task of considerable commercial interest as it enables musically intelligent training software. 
        
        Performance assessment is a ubiquitous aspect of music pedagogy: regular feedback from teachers improves the students' playing and auditions are used to place students in ensembles. It is, however, seen as a highly subjective and aesthetically challenging task with considerable disagreement on assessments between educators \cite{thompson_evaluating_2003,wesolowski_examining_2016}.
        An automatic system for assessment of performances would provide objective, reliable, and repeatable feedback to the student during practice sessions and increase accessibility of affordable instrument education. Generally, the structure of a performance assessment system resembles the basic structure of an audio content analysis system: features describing the performance are extracted and then used in the inference stage to estimate one or more ratings. 
        
        %From ISMIR paper
        %In some approaches, standard spectral and temporal features such as Spectral Centroid, Spectral Flux, and Zero-Crossing Rate were used \cite{knight_potential_2011}. In others, features aimed at capturing cognitive aspects of music perception were hand-designed using either musical intuition or expert knowledge \cite{abeser_automatic_2014,li_analysis_2015,nakano_automatic_2006,romani_picas_real-time_2015}. For instance, Nakano et al.\ used features measuring pitch stability and vibrato as inputs to a simple classifier to rate the quality of vocal performances \cite{nakano_automatic_2006}. Several studies also attempted to combine low-level audio features with hand-designed feature sets \cite{luo_detection_2015,vidwans_objective_2017,wu_towards_2016}, as well as incorporating information from the musical score into feature computation \cite{mayor_performance_2009,bozkurt_dataset_2017,devaney_automatically_2011,vidwans_objective_2017}. What about tempo and dynamics?
        The features might be simple standard features as used in other content analysis systems \cite{knight_potential_2011} or designed specifically for the task of performance assessment (e.g., pitch and timing stability) \cite{nakano_automatic_2006,abeser_automatic_2014,%li_analysis_2015,luo_detection_2015,
        wu_towards_2016}. Some systems also incorporate musical score information into the feature computation \cite{%mayor_performance_2009,
        devaney_automatically_2011,bozkurt_dataset_2017,vidwans_objective_2017}.
        
        The general trend in content analysis from feature design towards feature learning can also be observed in studies on performance assessment \cite{lerch_music_2019}, albeit a bit more reluctant. One of the reasons for this reluctance is the non-interpretability of learned features; an educational setting requires not only an accurate assessment but also an explanation of the reasons for that assessment (and possibly a strategy to improve the performance).
        
        The success rate of tools for automatic performance assessment still leaves room for improvement. Most of the presented systems either work well only for very select data \cite{knight_potential_2011} or have comparably low prediction accuracies \cite{vidwans_objective_2017,wu_towards_2016}, rendering them unusable in most practical scenarios.

\section{Music identification and categorization}
\label{sec:classification}
    A significant part of research in the area of ACA is concerned with the categorization of audio data, e.g., into musical genres. This group of tasks is~---from a technical point of view---~related to the estimation of similarity between different music files as well as the estimation of the emotion in a recording of music. Before we go into details on how to automatically categorize music, we will first look into music identification through audio fingerprinting, which was the first MIR technology with large scale adoption by consumers in the early 2000s.

    \subsection{Audio fingerprinting}
    \label{ssec:fingerprinting}
        \textit{Audio fingerprinting}\index{Audio Fingerprinting} aims at representing a recording in a small and unique so-called fingerprint (also: perceptual hash) in order to look up this recording in a previously prepared database and map it to the stored meta data. In contrast to many other presented systems in this chapter, fingerprinting is not concerned with extracting musical meaning from audio but solely with identifying a recording unambiguously. The two main applications of audio fingerprinting are
        \begin{inparaenum}[(i)]
            \item   the automated monitoring of broadcast stations for independent supervision of the broadcasted songs in order to verify broadcasting rights, and
            \item   end consumer services allowing the identification of audio in order to provide meta data such as artist name, song title, or album art \cite{cano_audio_2005}.
        \end{inparaenum}
        The design of an audio fingerprinting system has to solve multiple inherent core problems \cite{cano_review_2005}. On the one hand, the fingerprint has to be small in size to be transmitted and searched for efficiently, on the other hand, it has to be unique in order to identify one specific song from a database of possibly millions of songs. Furthermore, it has to be robust against quality degradations in the audio signal, as a user might record the audio of a song playing over speakers in a noisy environment. Last but not least, the fingerprint extraction has to be efficient enough to run on embedded and mobile devices.
        
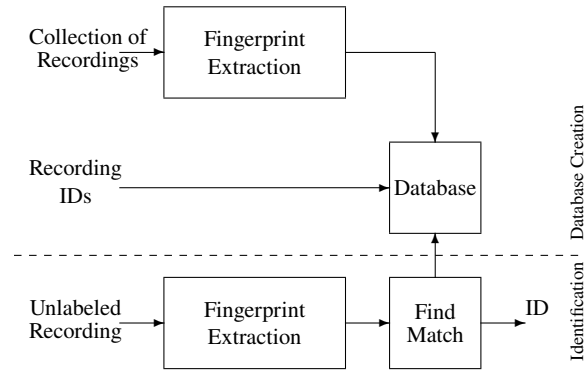
\begin{figure}
			\centering
			%\begin{footnotesize}
    \begin{picture}(76,50)
        \setcounter{iXOffset}{0}
        \setcounter{iYOffset}{0}
        \setcounter{iXBlockSize}{24}
        \setcounter{iXBlockSizeDiv2}{12}
        \setcounter{iYBlockSize}{12}
        \setcounter{iYBlockSizeDiv2}{6}
        \setcounter{iDistance}{6}

        \addtocounter{iYOffset}{\value{iYBlockSizeDiv2}}
        \addtocounter{iYOffset}{-2}

        \addtocounter{iXOffset}{2}
        \put(\value{iXOffset}, \value{iYOffset})
            {\text{{\shortstack[c]{Unlabeled\\ Recording}}}}
        \addtocounter{iXOffset}{\value{iDistance}}

        \addtocounter{iYOffset}{2}
        \addtocounter{iXOffset}{\value{iDistance}}

        \put(\value{iXOffset}, \value{iYOffset})
            {\vector(1,0){\value{iDistance}}}

        \addtocounter{iXOffset}{\value{iDistance}}
        \addtocounter{iYOffset}{-\value{iYBlockSizeDiv2}}
        
        \put(\value{iXOffset}, \value{iYOffset})
            {\framebox(\value{iXBlockSize}, \value{iYBlockSize}) {{\shortstack[c]{Fingerprint\\ Extraction}}}}

        \addtocounter{iXOffset}{\value{iXBlockSize}}
        \addtocounter{iYOffset}{\value{iYBlockSizeDiv2}}

        \put(\value{iXOffset}, \value{iYOffset})
            {\vector(1,0){\value{iDistance}}}

        \addtocounter{iXOffset}{\value{iDistance}}
        \addtocounter{iYOffset}{-\value{iYBlockSizeDiv2}}

        \put(\value{iXOffset}, \value{iYOffset})
            {\framebox(\value{iXBlockSizeDiv2}, \value{iYBlockSize}) {{\shortstack[c]{Find\\ Match}}}}

        \addtocounter{iXOffset}{\value{iXBlockSizeDiv2}}
        \addtocounter{iYOffset}{\value{iYBlockSizeDiv2}}

        \put(\value{iXOffset}, \value{iYOffset})
            {\vector(1,0){\value{iDistance}}}

        \addtocounter{iXOffset}{\value{iDistance}}
        \addtocounter{iYOffset}{1}

        %\addtocounter{iXOffset}{1}
        \put(\value{iXOffset}, \value{iYOffset})
            {\text{{\shortstack[c]{ID}}}}

        \setcounter{iXOffset}{0}
        \setcounter{iYOffset}{15}

        \multiput(\value{iXOffset}, \value{iYOffset})(2,0){38}
            {\line(1,0){1}}

        \addtocounter{iXOffset}{\value{iDistance}}
        \addtocounter{iXOffset}{\value{iDistance}}
        \addtocounter{iXOffset}{\value{iDistance}}
        \addtocounter{iXOffset}{\value{iDistance}}
        \addtocounter{iXOffset}{\value{iDistance}}
        \addtocounter{iXOffset}{\value{iDistance}}
        \addtocounter{iXOffset}{\value{iDistance}}
        \addtocounter{iXOffset}{\value{iDistance}}
        \addtocounter{iXOffset}{\value{iDistance}}
        \addtocounter{iXOffset}{\value{iDistance}}
        \addtocounter{iXOffset}{\value{iDistance}}
        \addtocounter{iXOffset}{\value{iDistance}}
        \addtocounter{iXOffset}{2}
        \addtocounter{iYOffset}{2}

        \put(\value{iXOffset}, \value{iYOffset})
            {\text{\rotatebox{90}{{\shortstack[c]{\scriptsize Database Creation}}}}}

        \setcounter{iYOffset}{1}
        \put(\value{iXOffset}, \value{iYOffset})
            {\text{\rotatebox{90}{{\shortstack[c]{\scriptsize Identification}}}}}

        \setcounter{iXOffset}{2}
        \setcounter{iYOffset}{18}

        \addtocounter{iYOffset}{\value{iYBlockSizeDiv2}}
        \addtocounter{iYOffset}{-2}

        \put(\value{iXOffset}, \value{iYOffset})
            {\text{{\shortstack[c]{Recording\\ IDs}}}}
        \addtocounter{iXOffset}{\value{iDistance}}
 
        \addtocounter{iYOffset}{2}
        \addtocounter{iXOffset}{\value{iDistance}}

        \put(\value{iXOffset}, \value{iYOffset})
            {\vector(1,0){36}}
        
        \addtocounter{iXOffset}{36}
        \addtocounter{iYOffset}{-\value{iYBlockSizeDiv2}}

        \put(\value{iXOffset}, \value{iYOffset})
            {\framebox(\value{iXBlockSizeDiv2}, \value{iYBlockSize}) {{\shortstack[c]{Database}}}}

        \addtocounter{iXOffset}{\value{iDistance}}
        \addtocounter{iYOffset}{-\value{iDistance}}
        \put(\value{iXOffset}, \value{iYOffset})
            {\vector(0,1){\value{iDistance}}}

        \setcounter{iXOffset}{2}
        \setcounter{iYOffset}{36}

        \addtocounter{iYOffset}{\value{iYBlockSizeDiv2}}
        \addtocounter{iYOffset}{-2}

        \put(\value{iXOffset}, \value{iYOffset})
            {\text{{\shortstack[c]{Collection of\\ Recordings}}}}
        \addtocounter{iXOffset}{\value{iDistance}}

        \addtocounter{iYOffset}{2}
        \addtocounter{iXOffset}{\value{iDistance}}

        \put(\value{iXOffset}, \value{iYOffset})
            {\vector(1,0){\value{iDistance}}}

        \addtocounter{iXOffset}{\value{iDistance}}
        \addtocounter{iYOffset}{-\value{iYBlockSizeDiv2}}
        
        \put(\value{iXOffset}, \value{iYOffset})
            {\framebox(\value{iXBlockSize}, \value{iYBlockSize}) {{\shortstack[c]{Fingerprint\\ Extraction}}}}

        \addtocounter{iXOffset}{\value{iXBlockSize}}
        \addtocounter{iYOffset}{\value{iYBlockSizeDiv2}}

        \put(\value{iXOffset}, \value{iYOffset})
            {\line(1,0){\value{iXBlockSizeDiv2}}}

        \addtocounter{iXOffset}{\value{iXBlockSizeDiv2}}
        \put(\value{iXOffset}, \value{iYOffset})
            {\vector(0,-1){\value{iXBlockSizeDiv2}}}

    \end{picture}
%\end{footnotesize}	
			\caption[Audio Fingerprinting]{Flowchart of a general fingerprinting system}
            \label{fig:FP_system}	
		\end{figure}

        A fingerprinting system has two main processing steps: the fingerprint extraction (often on a mobile devices at the client side) and the identification in a database (usually  on the server side), see Fig.~\ref{fig:FP_system} bottom. It can only identify recordings which were used for the database creation, see Fig.~\ref{fig:FP_system} top.
        
        \subsubsection{Fingerprint extraction}
        \label{ssec:fpextract}
            As the goal of audio fingerprinting is the identification of a specific recording (not a song, meaning it is supposed to differentiate between, e.g., a studio and a live version of the same song), the fingerprint does not have to contain musical information but can focus on the raw content of the audio. A simple predecessor of modern fingerprinting systems proposed for the identification of commercials in broadcast stems used  segments of the time domain envelope as a fingerprint \cite{lourens_detection_1990}. Nowadays, the fingerprint is usually derived from the STFT.
            There exist several approaches for fingerprint extraction. Two prominent approaches are 
            \begin{inparaenum}[(i)]
                \item   to encode spectral bands, more specifically the changes of band energy over both time and frequency in a binary format \cite{haitsma_highly_2002}, and
                \item   to identify salient peaks of the spectrogram and encode their relative location to each other \cite{wang_industrial_2003}.
            \end{inparaenum}
            The resulting size of the extracted fingerprint depends on the system: the first system, for example, represents three seconds of audio with \unit[8]{Kbit} \cite{haitsma_highly_2002}.
            
        \subsubsection{Fingerprint identification}
        \label{ssec:fpident}
            After successful extraction, the fingerprint of an unknown query signal has to be compared with a large number of previously extracted fingerprints in a database. Since this database can be large, this comparison has to be as efficient as possible in order to minimize processing time. Fingerprint systems use multiple tricks to speed up the lookup process, including hash lookup tables in which database entries are referenced by their fingerprint content or the use of fast-to-compute distance measures such as the Hamming distance \cite{hamming_error_1950}. Reordering the database entries according to their popularity can also decrease the average lookup time significantly.
            
    \subsection{Music genre classification}
    \label{ssec:genre}
        %this is also partly based on score information, but it might involve performance information and is only implicitly in the score
        Music Genre Classification (MGC)\index{Music Genre Classification} is a classical machine learning task that used to be one of the most popular tasks in the field of MIR. It is obviously useful to describe and categorize large collections of music to enable music discovery or content-based music recommendation systems. Despite its initial popularity and clear demand from users for classification systems, the task has some fundamental inherent problems that concern the subjective, inconsistent, and somewhat arbitrary definition of music genre. For example, there is disagreement on genre taxonomies and what they are based on (geography: ``Indian music, '' instrumentation: ``symphonic music,'' epoch: ``baroque,'' technical requirements: ``barbershop,'' or use of the music: ``Christmas songs'') \cite{pachet_taxonomy_2000}. Despite these issues, MGC systems can be ``successfully'' developed as long as they adhere to one consistent and coherent definition of genre. The performance of the system strongly depends on
        \begin{inparaenum}[(i)]
            \item   the features and the information they are able to transport,
            \item   the data that the machine learning system is being trained on, and
            \item   the capabilities of the machine learning system or classifier itself.
        \end{inparaenum}
        
        \subsubsection{Feature examples}
        \label{ssec:features}
            Even as the definition of genre is problematic as pointed out above, it is obvious to the observant listener that the identification of specific genres may require information related to timbre, pitch, rhythm and tempo, and dynamics. Traditional systems with custom-designed features therefore use a variety of different features. Hundreds and possibly thousands of different features\index{Features} have been investigated over the years for their performance in MGC systems. The most successful features have been shown, interestingly enough, to be very simple features describing the spectral shape and the intensity of the signal. 
            
            Many of these well-known low-level features are extracted from short blocks of audio samples, resulting in a time series of values for each feature. In a next step, features are often aggregated over time by computing, for example, their mean or their standard deviation. This means that a complete audio file is ultimately represented by one vector with a length of one or two (mean and standard deviation) times the number of features.
            
            \paragraph{Spectral shape (timbre) features}
                Features describing the spectral shape of a signal are widely used. The spectral shape significantly influences our perception of the timbre of a sound \cite{helmholtz_lehre_1870}. Most of these features are extracted from the local magnitude spectrum (one column of the spectrogram as shown in Fig.~\ref{fig:specgram}).
                Here, only two commonly used features which have proven useful in analysis tasks will be presented as representatives of a large number of common features.
                \begin{itemize}
                    \item   \textit{Spectral Centroid}: The spectral centroid\index{Spectral Centroid} is the center of mass of a magnitude spectrum, i.e., the frequency at which the spectral magnitudes can be separated into two equal parts. That means that low frequency signals will have a low centroid while substantial high frequency components and noise will increase the centroid. Despite its technical definition, the spectral centroid has been shown to be strongly correlated to the perceptual sound attribute brightness \cite{mcadams_perceptual_1995,caclin_acoustic_2005}.
                    %\item   \textit{Spectral Rolloff}: The Spectral Rolloff is the frequency below which the majority of the spectral energy is found. That means that the more high frequencies are in the spectrum, the higher the Rolloff will be. Thus, the Spectral Rolloff is a measure of signal bandwidth.
                    \item   \textit{Mel Frequency Cepstral Coefficients}: The Mel Frequency Cepstral Coefficients (MFCCs)\index{Mel Frequency Cepstral Coefficients} are a measure of the cosine-shape of the STFT on a logarithmic (Mel-shaped) frequency axis. To compute, the STFT axis is first split into logarithmically spaced frequency bands according to the Mel-scale modeling the human frequency perception \cite{stevens_scale_1937}. Then, a logarithm is applied to the spectrum and the bands are transformed into the cepstral domain with a Discrete Cosine Transform (DCT). The resulting DCT coefficients are the MFCCs. In contrast to the  spectral centroid, the MFCCs are a multidimensional feature similar to the pitch chroma (Sec.~\ref{ssec:chroma}), often representing each spectrum with $13$ or more values.
                    
                    The algorithm for the MFCC computation is an interesting mixture of ideas from perception, signal processing (cepstrum), and data compression (DCT) \cite{davis_comparison_1980}.
                    While the interpretability of the MFCCs is limited and their perceptual meaning is questionable, they have proven surprisingly useful in a wide variety of tasks \cite{logan_mel_2000,jensen_evaluation_2006,heittola_musical_2009} and are nowadays considered a standard baseline feature. % \cite{}.
                \end{itemize}
            
            \paragraph{Intensity features}
                Intensity features model the dynamics or loudness of a recording. The two feature examples given below represent two simple and common features.
                \begin{itemize}
                    \item   \textit{Envelope}: The simplest way to describe the envelope of a signal per block is to find the absolute maximum per block, resulting in the overall shape of the waveform. This is somewhat related to the slightly more complicated Peak Programme Meter (PPM) that is used in recording studios.
                    \item   \textit{Root Mean Square}: The Root Mean Square (RMS)\index{Root Mean Square} is a standard way of computing the intensity of a signal. It is the so-called effective value of the signal computed as the square root of the mean of the squared values per block. For long block sizes, it can be an efficient measure of long-term level changes. A common pre-processing step is to filter the signal before computing the RMS to take into account the sensitivity of the human ear at different frequencies for different level ranges (compare A-weighted or C-weighted sound pressure level measures).
                \end{itemize}
            
            \paragraph{Additional features}
                Musical genre is so broadly defined that features representing characteristics from many categories can be meaningful. Therefore, numerous and diverse custom-designed features have been investigated for this task. These features include, for example,
                \begin{inparaitem}[]
                    \item   stereo features representing the width and form of the stereo image \cite{tzanetakis_stereo_2007},
                    \item   pitch content features representing the variety and ranges of pitches \cite{tzanetakis_pitch_2002}, and
                    \item   tempo and rhythm features describing tempo variation, beat strength, and rhythmic complexity \cite{dixon_classification_2003,burred_hierarchical_2004}.
                \end{inparaitem}
                Many of these features are, unlike the features introduced above, extracted from longer windows of data. For instance, attempts of describing the rhythmic content need a context window of approx.\ \unit[5]{s} or more to be meaningful.
                As pointed out above, the era of feature design is nowadays considered a thing of the past (except for problems with only small amounts of available data) and modern features are often learned directly from the spectrogram. Common feature learning approaches are based, for example, on neural networks \cite{lee_unsupervised_2009,hamel_learning_2010} or dictionary learning \cite{mairal_online_2009, wu_assessment_2018}.
            
        %\subsubsection{Data}
        %\label{ssec:data}
            %A machine learning system is a data-driven system, meaning that it learns the most likely mapping between input (features) and output (classes) from a set of data. This means that in order to train a system well, there exist a few important requirements on data.
            %\begin{inparaitem}[]
                %\item   First, the training data have to be representative. That means that the possible variability of the input data should be covered completely to enable the system to learn. A music genre can, for instance, not be properly represented by only one band as the system might learn to distinguish the band but not the genre. If the number of songs per class is imbalanced, then this should match the distribution of the unseen inputs that the system will be used for later.
                %\item   Second, the training data have to be sufficient. The more complex a task is, the more complex a system needs to be, and the more complex a machine learning system is, the more training data is required. Without a sufficient amount of data, a complex system will either not learn a proper mapping or it will ``overfit, '' meaning that the system works very well on the data is has been trained on but poorly on unseen data.
                %\item   Third, the training data should not be noisy, meaning the labels should be consistent and unambiguous. This is, as mentioned above, a problem for MGC because of the issues outlines with the definition of music genre above.
            %\end{inparaitem}
            
        \subsubsection{Classification}
        \label{ssec:classifier}
\begin{figure}%
            \centering
            \includegraphics[width=.7\columnwidth]{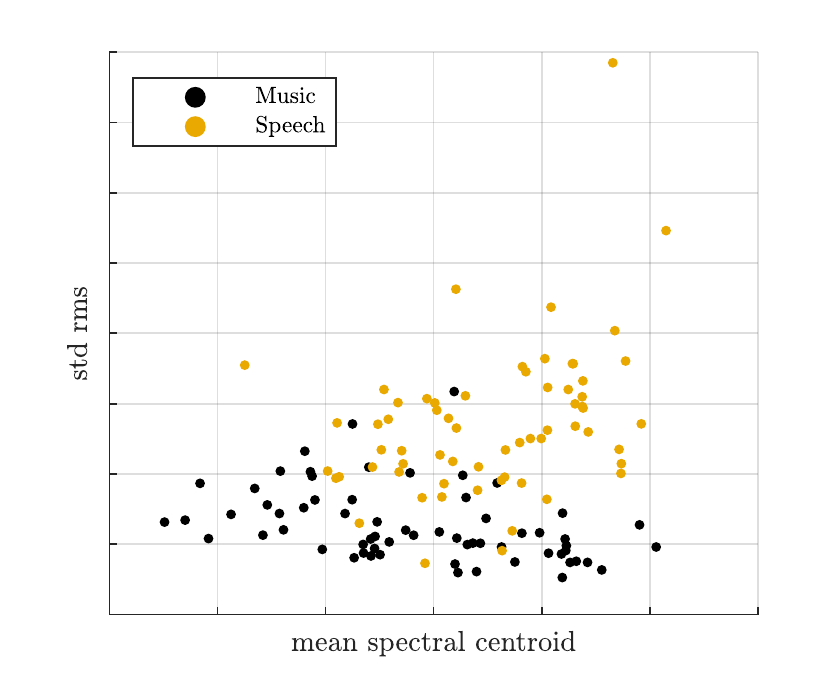}%
            \caption{A music/speech dataset visualized in a two-dimensional feature space (x-axis: average spectral centroid, y-axis: standard deviation of rms)}%
            \label{fig:featurespace}%
            \end{figure}

            The simplest, most intuitive classifier is just a threshold: if a feature value is higher than a threshold $\epsilon$ choose class 1, otherwise class 2. A modern data-driven system derives this threshold from the data itself and generalizes to a multi-dimensional space with possibly non-linear thresholds. In other words, it learns what combination of feature values are common for each class and how to differentiate between classes given these feature values. Figure~\ref{fig:featurespace} shows a so-called scatter plot for two classes represented by two features (two-dimensional feature space). It can be seen that for this data, the RMS feature seems to work slightly better in separating the two classes speech and music than the spectral centroid. This visualization also emphasizes the importance of the so-called training data set; if the estimated thresholds are based on data that are not representative, the classifier will not perform well.
            
            Another example for a basic classifier is the Nearest-Neighbor classifier \cite{fix_discriminatory_1951}. While training, it stores the location of each data point in the feature space with its corresponding class label. When queried with a new and unseen feature vector, the distance to every single training vector is computed; the final result is the class label of the closest vector, the nearest neighbor.
            Other classifiers model the distribution of data, avoiding to store each individual data point and allowing for simple generalization of the data. One popular classifier that models the data with Gaussian distributions is the Gaussian Mixture Model \cite{duda_pattern_2000}. 
            More modern approaches maximize the separation between classes by mapping the data into high-dimensional spaces (Support Vector Machine \cite{boser_training_1992}) or form a so-called ensemble of many simple classifiers to yield a more robust majority vote (Random Forest \cite{breiman_random_2001}). Most state-of-the-art classifiers are, if the amount of training data allows it, based on (deep) neural networks \cite{goodfellow_deep_2016}.

    \subsection{Music emotion recognition}
    \label{ssec:emotion}
        An analysis task for which many consumers seem to see an intuitive need is the extraction of the emotional content of a recording, as a majority of users search for and categorize music by describing the emotional content of the music \cite{kim_categories_2002}. The task definition of \textit{Music Emotion Recognition} (MER)\index{Music Emotion Recognition} suffers from similar, possibly more severe, restrictions in terms of subjectivity and noisiness of human-labeled data. The main issues are
        \begin{inparaenum}[(i)]
            \item   the question of whether to estimate the conveyed emotion or the elicited emotion, i.e., a subject might perceive the music transporting a specific emotion, but might feel a completely different emotion themselves  \cite{meyer_emotion_1956,zentner_emotions_2008}, and
            \item   the unclear definition of what emotions or moods are actually inherent to music listening, e.g., can music trigger basic emotions such as fear and anger \cite{scherer_why_2003, scherer_which_2004,zentner_emotions_2008}?
        \end{inparaenum}
        Due to the (commercially) appealing applications of MER, however, these inherent problems have not stopped researchers from targeting this task \cite{yang_machine_2012}. Despite many studies investigating emotional content and impact of music, however, the link between musical parameters to emotions remains largely unknown, and the audio features which could directly describe emotional content are unknown as well. Therefore, the features and approaches commonly chosen for MER are similar to genre classification as described in Sect.~\ref{ssec:genre}. In addition to such classification approaches, some methods aim at estimating emotions not by sorting them into distinct classes but locating them as coordinates in a two-dimensional valence-arousal plane as proposed by Russel \cite{russel_circumplex_1980}. In this case, the machine learning system is not choosing one of multiple output categories but estimates two values (valence and arousal). This class of machine learning algorithms is referred to as regression algorithms.
        
    %\subsection{Summary}
    %\label{ssec:summary}

\section{Current challenges}
\label{sec:challenges}
    Looking at the historic development in audio content analysis as well as the currently pressing research issues, three main overall challenges can be identified, 
    \begin{inparaenum}[(i)]
        \item   the amount of data available for training complex machine learning systems, 
        \item   the predictability of modern systems and interpretability of the results, and
        \item   the inherently abstract musical language and the largely unclear relations of musical concepts and perceptual meaning.
    \end{inparaenum}
    
    \subsection{Training data}
    \label{ssec:train}
        Machine learning systems are data-driven, meaning that they learn the most likely mapping between input (features) and output (classes) from a set of data. Thus, in order to train a system well, there exist important requirements on data.
        \begin{inparaitem}[]
            \item   First, the training data have to be representative. That means that, on the one hand, the possible variability of the input data should be covered completely to enable the system to learn. A music genre can, for instance, not be properly represented by only one band as the system might learn to distinguish the band but not the genre. On the other hand, the distribution of songs per class should reflect the expected distribution so as to not bias the system towards majority classes.
            \item   Second, the training data should not be noisy, meaning the labels should be consistent and unambiguous. This is, as mentioned above, a problem especially for MGC because of the issues with the definition of music genre.
            \item   Third, the training data have to be sufficient. The more complex a task is, the more complex a system needs to be, and the more training data is required. Without a sufficient amount of data, a complex system will not be able to generalize a model from the training data and thus will ``overfit,'' meaning that the system works very well on the data it has been trained on but poorly on unseen data \cite{duda_pattern_2000}.
        \end{inparaitem}
        The amount of training data becomes a crucial issue for machine learning approaches and systems based on Deep Neural Networks which have shown superior performance at nearly all tasks in audio content analysis but require large amounts of data for training. 
    
        Although there is a vast amount of music data easily accessible, not all of these data can be directly used for training a machine learning system. A system for transcribing drum events from popular music, for instance, needs expert annotations precisely marking each drum hit in terms of instrument and timing (and possibly the playing technique). Marking these individual hits is, however, a very time-consuming and tedious task so that the increasing requirement for data due to the increasing system complexity usually outpaces the annotation of new data by human annotators. Given the multitude of annotations needed for various content analysis tasks, it is likely that the gap between available annotated data and required amount of training data will widen with increasing system complexity. This will result in a growing need of systems and approaches addressing this challenge, as is reflected by an increasing research interest in this problem from various angles. Current approaches include:
        \begin{itemize}
            \item   \textit{data augmentation and synthetic data}: without sufficient annotated data, the machine learning engineer can ``cheat'' by virtually increasing the amount of training data either with synthesized data (e.g., through synthesis of MIDI data) or by processing existing audio data with irrelevant transformations, for instance, pitch shifting for music genre classification or segmenting the longer file into shorter segments \cite{mignot_analysis_2019};
            \item   \textit{transfer learning}: although data might be scarce for one task, it might be available for related tasks; therefore, the idea of transfer learning is to take an internal representation of a system trained for one task with abundant data and use this (hopefully powerful) representation as a feature input to a more simple classifier that can be trained with significantly less data \cite{choi_transfer_2017};
            \item   \textit{weakly labeled data}: annotating audio data with high accuracy is a tedious and time-consuming task, however, it becomes significantly less demanding if high time accuracy is not required by, e.g., labeling the presence of a musical instrument in a snippet of audio without pinpointing its exact occurrence time(s); this requires, however, to modify existing machine learning approaches to deal with the weak time accuracy \cite{gururani_attention_2019};
            \item   \textit{self-supervised systems and unsupervised systems} which tackle the data challenge in MIR in a different way by exploring possibilities of training systems with unlabeled data; it is, for example, possible to train a complex system with high data requirements with unlabeled data by utilizing the outputs of pre-existing systems for these unlabeled data as training targets for the new system \cite{wu_labeled_2018} or by utilizing synthesis iteratively to learn from unlabeled data \cite{choi_deep_2019}.
                %\begin{itemize}
                    %\item   generative
                    %\item   student-teacher
                    %\item   keuwnoo \cite{ismir19}
                %\end{itemize}
        \end{itemize}
        As the field of machine learning evolves rapidly, it is difficult to predict if any of the methods above will become standard approaches. It is clear, however, that the lack of large-scale training data will continue to impact the progress and methods applied to audio content analysis.
        
    \subsection{Interpretability and Understandability}
    \label{ssec:interpretability}
        The increasing complexity of machine learning systems not only affects the required amount of training data, but also leads to highly complex systems which do not allow easy interpretation of internal states, easy analysis of results, or understanding of influencing factors \cite{lipton_mythos_2018}. In traditional feature-driven designs, each feature had at the minimum a technical connotation allowing an expert to connect system outputs to somewhat interpretable inputs. As modern systems learn features from the data, it becomes increasingly difficult to derive this link, thus limiting the interpretability and control over system behavior. This restricts possibilities to tweak system outputs for specific inputs and could increase the likelihood of the system producing unexpected results especially for unseen data. Recent years have seen this problem partly being addressed in the field of audio content analysis through
        \begin{inparaenum}[(i)]
            \item   visualization that gives insights into the networks' internal states or intermediate representations, diagnoses the embedded space, or disentangles the complex internal representations into interpretable graphs \cite{zhang_visual_2018},
            \item   enforcing interpretable latent spaces or intermediate representations that are humanly understandable through regularization \cite{hadjeres_glsr-vae_2017,brunner_midi-vae_2018}, and
            \item   analysis and transformation of latent spaces to interpretable spaces \cite{adel_discovering_2018}.
        \end{inparaenum}

    \subsection{Perceptual meaning}
    \label{ssec:preception}
        A challenge unrelated to the technical progress but somewhat emphasized by the engineering-driven methodologies in the field is the question of the perceptual relevance of the results of ACA-systems. The problems of data annotation in the context of musical genre and emotion have already been discussed in Sects.~\ref{ssec:genre} and \ref{ssec:emotion}, however, there are other fundamental issues when it comes to ``musical meaning.'' While, for example, the task of extracting drum onset times from audio is clearly defined, it is less clear what higher musical meaning can be derived from it. It is not easy to find answers to questions such as
        \begin{inparaitem}[]
            \item   is there a generalized way of describing rhythm and rhythmic properties, and
            \item   can specific rhythmic properties and/or timing variations be mapped to specific affectual responses in humans?
        \end{inparaitem}    
        A major obstacle impeding MIR research is the inability to successfully isolate (and therefore understand) the various score-based and performance characteristics that contribute to the music listening experience. The listener, however, has to be the ultimate judge of the usefulness of any audio content analysis system \cite{lerch_music_2019}.

\section{Outlook}
\label{sec:outlook}
    Audio Content Analysis is an emerging research field facing interesting challenges and enabling a wide range of future applications. 
    %While the previous sections have mostly introduced examples for analysis tasks in the field, the range of possible applications, both current and future, is wide. 
    We are already seeing new applications and emerging companies building on the advances made in the field.
    
    \subsection{Music education}

    %APPLIED SCIENCES
    The idea of utilizing technology to assist music (performance) education is not new. Seashore pointed out the value of scientific observation of music performances for improving learning as early as the 1930s \cite{seashore_psychology_1938}. One of the earliest studies exploring the potential of computer-assisted techniques in the music classroom was carried out by Allvin \cite{allvin_computer-assisted_1971}. Although his work was focused on using technology for providing individualized instruction and developing aural and visual aids to facilitate learning, it also highlighted the potential of using ACA techniques such as pitch detection to perform error analysis in a musical performance and provide constructive feedback to the learners through (semi-)automated music tutoring software.  Such software aims at supplementing %or replace \ashis{\pcount Ashis: do we want to say replace ?} 
		teachers by providing students with insights and interactive feedback by analyzing and assessing the audio of practice sessions. The ultimate goals of an interactive music tutor are to highlight problematic parts of the students' performance, provide a concise yet easily understandable analysis, give specific and understandable feedback on how to improve, and individualize the curriculum depending on the students' mistakes and general progress.
    
    %MPA
    %Over the last decade, several researchers have worked towards developing tools capable of automatic music performance assessment which can be categorized based on
    %\begin{inparaenum}[(i)] 
        %\item the parameters of the performance that are assessed, and
        %\item the technique/method used to design these systems.
    %\end{inparaenum}
    
    Tools for performance assessment typically assess one or more performance parameters which are usually related to the accuracy of the performance in terms of pitch and timing \cite{wu_towards_2016,vidwans_objective_2017,pati_assessment_2018,luo_detection_2015} or quality of sound (timbre) \cite{knight_potential_2011,romani_picas_real-time_2015}. Various (commercial) solutions are already available, exhibiting a similar set of goals. These systems adopt different approaches, ranging from traditional music classroom settings to games targeting a playful learning experience. Examples for tutoring applications are SmartMusic\footnote{{MakeMusic, Inc.}, SmartMusic, \url{https://www.smartmusic.com}, last accessed 04/11/2019}, Yousician\footnote{{Yousician Oy}, Yousician, \url{https://www.yousician.com}, last accessed 04/11/2019}, Music Prodigy\footnote{The Way of H, Inc. (dba Music Prodigy), {Music {Prodigy}}, \url{http://www.musicprodigy.com}, last accessed 04/11/2019}, and SingStar\footnote{Sony Interactive Entertainment, {SingStar}, \url{http://www.singstar.com}, last accessed 04/11/2019}. 
    
        %The most obvious application example connecting MPA and MIR is music tutoring software.
   
    \subsection{Music production}
        Knowledge of the audio content enables the improvement of music production tools in various dimensions. 
        The most obvious enhancement can be found in terms of productivity and efficiency: the better a software understands the details of incoming audio streams or files, the better it can adapt, for instance, by applying default gain and equalization parameters \cite{reiss_applications_2018} or suggest compatible recordings from a library. Systems might support editors by automatic artifact-free splicing of multiple recordings from one session or selecting error-free recordings from a set of recordings. Modern processing methods allow for subtle or dramatic timing and pitch variations in high quality\footnote{Antares, Autotune, \url{https://autotune.com}, last accessed 01/14/2020}\footnote{zplane elastique, \url{https://licensing.zplane.de/technology#elastique}, last accessed 01/14/2020}~---~controlling them with musically relevant content-adaptive intelligence could streamline music production in unprecedented ways.
        
        Modern tools also enhance the creative possibilities in the production process. For example, creating harmonically meaningful background choirs by analyzing the lead vocals and the harmony track is already technically feasible nowadays.\footnote{zplane vielklang, \url{https://vielklang.zplane.de}, last accessed 01/14/2020} Knowing and possibly separating sound sources in a recording could enable new ways of modifying or morphing different sounds to create new soundscapes and auditory scenes. 
        Many more scenarios are conceivable where audio analysis will impact the production process, although the multifaceted character of the field makes it difficult to predict specific use cases.

    \subsection{Music distribution and consumption}
        Audio analysis has already started to transform consumer-facing industries such as streaming services with audio-based music recommendation and playlist generation systems using an in-depth understanding of the musical content \cite{knees_intelligent_2019}. This is not only the case for the end-consumer themselves: there is also a industry need for automatically identifying music and creating playlists that conform to the company's brand image \cite{herzog_predicting_2017}.
        %The trend towards parametric and object-oriented audio coding (see Chap.~\ref{X}) continues to improve audio quality at low bitrates significantly. It requires, however, a detailed understanding of the sources and the auditory scene of the audio signal to be encoded. The more details are known about the content, the more efficiently it can be encoded and transmitted.
        %
        %One of the most obvious end consumer applications of content analysis is in music discovery tools, automatic music recommendation systems, as well as playlist generation systems as most of them depend on a comprehensive understanding of the musical content \cite{knees_intelligent_2019}.
        
        In the near future, we can expect the rise of creative music discovery and listening applications that enable the listener to interact not only by choosing content but interact with the content itself. This could include, for example, the gain adjustment for individual voices, replacing instruments or vocalists, or interactively changing the musical arrangement.
        
    \subsection{Generative music}
        An important outcome of being able to extract machine interpretable content information from audio data is the possibility for these data to feed generative algorithms. The automatic composition and rendition of music is emerging as a  challenging yet popular research direction \cite{briot_deep_2020}, gaining interest from both research institutions and industry. While bigger questions concerning capabilities and restrictions of computational creativity as well as aesthetic evaluation of algorithmically generated music remain largely unanswered, practical applications such as generating background music for user videos and commercial advertisements are currently in the focus of many researchers. The interactive and adaptive generation of sound tracks for video games as well as individualized generation of license-free music content for streaming are additional long-term goals of considerable commercial interest.

\bibliographystyle{spmpsci}
\bibliography{2019-hda-aca}

\printindex

\end{document}